\documentclass[12pt]{article}

\newcommand{\br}{{\bf {R}}}

\newcommand{\Om}{\Omega}
\newcommand{\om}{\omega}
\newcommand{\Tr}{\rm{Tr}}
\newcommand{\tr}{\rm{Tr}}

\newcommand{\bs}{{\bf S}}

\usepackage{latexsym,amsfonts,amsmath,amsthm,amssymb}

\usepackage[T1]{fontenc}
\usepackage[latin1]{inputenc}
\title{Prequantum classical statistical model with infinite dimensional phase-space-2: complex representation of symplectic phase-space model}
\author{Andrei Khrennikov\\
School of Mathematics and Systems Engineering\\
University of V\"axj\"o, S-35195, Sweden}

\begin{document}
\maketitle

\abstract{We show that QM can be represented as a natural projection of a classical statistical model on the phase space
$\Omega= H\times H,$ where $H$ is the real Hilbert space. Statistical states are given by Gaussian measures on $\Omega$
having zero mean value and dispersion of the Planck magnitude -- fluctuations of the ``vacuum field.'' Physical variables (e.g., energy) are given by maps $f: \Omega \to {\bf R}$ (functions of classical fields).  The crucial point is that statistical states and variables are symplectically invariant. The conventional quantum representation of our prequantum classical statistical model is constructed on the basis of the Teylor expansion (up to the terms of the second order  at the vacuum field point $\omega=0)$ of  variables $f: \Omega \to {\bf R}$ with respect to the small parameter $\kappa= \sqrt{h}.$ A Gaussian symplectically invariant measure (statistical state) is represented by its covariation operator (von Neumann statistical operator).
A symplectically invariant smooth function (variable) is represented by its second derivative at the vacuum field point $\omega=0.$ From the statistical viewpoint QM is a statistical approximation of the prequantum classical statistical field theory (PCSFT). Such an approximation is obtained through neglecting by statistical fluctuations of the magnitude $o(h), h\to 0,$ in averages of physical variables. Equations of Schr\"odinger, Heisenberg and von Neumann are  images of dynamics on $\Omega$ with a symplectically invariant Hamilton function.}

\section{Introduction}
In the first part of this paper [1] we demonstrated that, in spite of all ``NO-GO'' theorems,
it is possible to construct a prequantum classical statistical model. The phase space of this model
is the infinite dimensional Hilbert space ; so classical ``systems'' are in fact classical fields.
We call this approach the {\it prequantum classical statistical field theory} (PCSFT).
There was constructed a natural map $T$ establishing the correspondence between classical and
quantum statistical models.
The cornerstone of our approach is that the correspondence map $T$ should approximately
preserve averages (up to fluctuations of the magnitude $o(h), h \to 0):$
\begin{equation}
\label{AVP}
<f>_\rho = <T(f)>_{T(\rho)} + o(h),
\end{equation}
where $\rho$ and $f$ are, respectively, classical statistical states and variables.
In particular, for the space of physical variables $V_{\rm{quad}}$ consisting of quadratic forms on the
Hilbert space, we have the precise equality of classical and quantum averages:
\begin{equation}
\label{AVP1}
<f>_\rho = <T(f)>_{T(\rho)}
\end{equation}
and the correspondence between classical variables and quantum observables is one-to-one. For the space of analytic
physical variables, we have only asymptotic equality (\ref{AVP}) and the correspondence between classical variables and quantum observables is not one-to-one. A huge class of classical variables is mapped into the same quantum observable.
In [1] there was chosen the space of statistical states $S_G^h$  consisting of Gaussian measures (with zero mean value) having dispersion equal to the Planck constant:
$$
\sigma^2(\rho) =\int \Vert \omega \Vert^2 d \rho (\omega)= h.
$$
Quantum states (pure as well as mixed) are  images of Gaussian fluctuations
of the magnitude $h$ on the infinite dimensional space.

In [1] we considered the quantum model based on the real Hilbert space $H.$ This model is essentially
simpler than the complex quantum mechanics, because clarification of introduction  of the complex structure
on the phase space is a very complicated problem. We analyse this problem in this article. We shall show that
the complex structure is nothing else than the image of the symplectic structure on the infinite dimensional
phase space. By using symplectic structure we find classes of classical physical variables and statistical states.
Here we should apply dynamical arguments. We found the classical
Hamiltonian dynamics on the phase space which induces the quantum state dynamics (Schr\"odinger's equation). The crucial point is that the classical Hamilton function ${\cal H}(\omega)$ should be {\it symplectically invariant:}
\begin{equation}
\label{SI}
{\cal H}(J\omega)= {\cal H}(\omega),
\end{equation}
where $\omega \in \Omega= Q\times P, \; Q=P=H,$ and $J: Q\times P \to Q\times P$ is the symplectic operator.
At the beginning we restrict our considerations to quadratic physical variables.
The  space of classical observables is chosen consisting of {\it symplectically invariant quadratic forms.}
In such a model the correspondence $T$ between classical variables and  quantum observables is one-to-one.

There is another motivation to consider classical dynamics with symplectically invariant Hamilton functions,
namely only such dynamics preserves the magnitude of classical random fluctuations: dispersion of a Gaussian
measure. The  space of classical statistical states is chosen consisting of {\it symplectically invariant
Gaussian measures} having dispersion  of the magnitude $h$ (and zero mean value).

We pay attention that any point wise classical dynamics (in particular, Hamiltonian) can be lifted to
spaces of variables (functions) and statistical states (probability measures).
In the case of symplectically invariant Hamilton function by mapping these liftings
to the quantum statistical model we obtain, respectively,  Heisenberg's
dynamics for quantum observables and von Neumann's dynamics for statistical operators.

We emphasize that
one could not identify classical point wise state dynamics and dynamics of statistical states. In the conventional quantum mechanics these dynamics are typically identified. Our approach supports the original views of
E. Schr\"odinger. His equation describes the evolution of classical states (fields). It is impossible to provide
any statistical interpretation to such states. In particular, wave function considered as a field satisfying
Schr\"odinger's equation has no statistical interpretation. Only statistical states (probability measures in the classical model) and corresponding density operators (which are in fact covariation operators of measures) have a statistical interpretation. The root of misunderstanding was assigning (by M. Born) the statistical interpretation to a normalized
wave function. The tricky thing is that in fact Born's interpretation should be assigned not to an individual state $\psi,$ but to a statistical state corresponding to the Gaussian distribution with the covariation operator:
\begin{equation}
\label{PSI}
B_\psi = h\psi \otimes \psi.
\end{equation}
Thus pure quantum states are simply statistical mixtures of special Gaussian fluctuations
(concentrated on two dimensional (real) subspaces of the infinite dimensional Hilbert space), see
section 9 for details. Of course, one could reproduce dynamics of such a statistical state by considering
the Schr\"odinger equation with random initial conditions:
\begin{equation}
\label{SRI}
i h \frac{d\xi}{d t}(t;\omega) = {\bf H} \xi(t;\omega), \xi(t_0;\omega)= \xi_0(\omega),
\end{equation}
where ${\bf H}$ is Hamiltonian and $\xi_0(\omega)$ is the initial Gaussian random vector taking values in  the Hilbert space. We emphasize that $\Vert \xi(t;\omega) \Vert\in [0, +\infty).$

\section{Symplectically invariant classical mechanics}

{\bf 2.1. Dynamics induced by a quadratic Hamilton function.} We consider the conventional classical phase space:
$$
\Om=Q \times P, Q= P = \br^n
$$
Here states are represented by points $\om=\{q, p\} \in \Om;$ evolution of a state is described by the Hamiltonian equations
\begin{equation}
\label{HE}
\dot q = \frac{\partial {\cal H}}{\partial p},\; \;  \dot p=-\frac{\partial {\cal H}}{\partial q,}
\end{equation}where ${\cal H}(q, p)$ is the Hamilton function (a real valued function on the phase space $\Om).$
We consider the scalar product on $\br^n:$
$
(x, y)=\sum_{j=1}^n x_j y_j
$
and define the scalar product on $\Omega:$
$(\om_1, \om_2)=(q_1, q_2) +(p_1, p_2).$
In our reseach we shall be interested in a quadratic Hamilton function:
\begin{equation}
\label{V}
{\cal H}(q, p)=\frac{1}{2} ({\bf H} \om,\om),
\end{equation}where ${\bf H}: \Om \to \Om$ is a symmetric operator.
We remark that any ($\br$-linear) operator $A: {\bf R}^{2n} \to {\bf R}^{2n}$ can be represented in the form
\[A= \left( \begin{array}{ll}
A_{11}&A_{12}\\
A_{21}&A_{22}
\end{array}
\right ),
\]
where $A_{11}: Q \to Q, A_{12}: P \to Q,$ $A_{21}: Q \to Q, A_{22}: P \to P.$
A linear operator $A: {\bf R}^{2n} \to {\bf R}^{2n}$ is symmetric if $$A_{11}^*=A_{11}, A_{22}^*=A_{22},
A_{12}^*=A_{21}, A_{21}^*=A_{12}.
$$
Thus the Hamilton function (\ref{V}) can be written as:
\begin{equation}
\label{V0}
{\cal H}(q,p) = \frac{1}{2} [({\bf H}_{11} q, q)+2({\bf H}_{12} p, q) + ({\bf H}_{22} p,p)],
\end{equation}
The Hamiltonian equation has the form:
\begin{equation}
\label{HE20}
\dot q= {\bf H}_{21}q + {\bf H}_{22} p,  \; \;   \dot p=-( {\bf H}_{11}q +{\bf H}_{12}p)
\end{equation}
As always,  we define the symplectic structure  on the phase space starting with the symplectic operator
 \[J= \left( \begin{array}{ll}
 0&1\\
 -1&0
 \end{array}
 \right ),
 \]
(here the blocks "$\pm 1$" denote $n \times n$ matrices with $\pm 1$ on the diagonal).
By using the symplectic operator $J$ we can write these Hamiltonian equations in the operator form:
\begin{equation}
\label{Y}
\dot \om= \left( \begin{array}{ll}
\dot q\\
\dot p
\end{array}
\right )=J{\bf H} \om
\end{equation}
or
\begin{equation}
\label{Y1}
-J \dot \om= {\bf H} \om
\end{equation}
From (\ref{Y}) we get
\begin{equation}
\label{Y2}
\om(t)= U_t \om, \; \; \mbox{where} \; U_t=e^{J {\bf H} t}.
\end{equation}
The map $U_t\omega$ is a linear Hamiltonian flow on the phase space $\Omega.$

{\bf 2.2. Symplectically invariant quadratic forms and \\ $s$-commuting operators.}
In our investigations we shall be concentrated on consideration of symplectically invariant
quadratic forms. It is easy to see that symplectic invariance of the quadratic form $f_A(\omega)=
(A\omega, \omega)$, where $A: \Omega \to \Omega$ is the linear symmetric operator,  is equivalent to commuting of $A$ with the symplectic operator $J.$

Let us consider the class ${\cal L}_{\rm symp} \equiv {\cal L}_{\rm symp} (\Omega)$ of (all) linear operators $A: \Omega \to \Omega$ which commute with the symplectic operator:
\begin{equation}
\label{SS}
A J= J A
\end{equation}

This is a subalgebra of the algebra of all linear operators ${\cal L}_{\rm symp} (\Omega).$

\medskip

{\bf Proposition 2.1.} {\it $A\in {\cal L}_{\rm symp}$ iff $A_{11}=A_{22}=D, A_{12}=-A_{21}=S$, i.e.,}
\[A= \left( \begin{array}{ll}
D&S\\
-S&D
\end{array}
\right )
\]

We remark that an operator $A \in {\cal L}_{\rm symp}$ is symmetric iff $D^*=D$ and $S^*=-S.$  Hence any symmetric $s$-commuting operator in the phase space is determined by a pair of operators $(D, S)$, where $D$ is symmetric and $S$ is anti-symmetric. Such an operator induces the quadratic form
\begin{equation}
\label{QF}
f_A(\om)=(A\om, \om)=(Dq, q)+2(Sp, q)+(Dp, p).
\end{equation}

{\bf 2.3. Dynamics for symplectically invariant quadratic Hamilton functions.}
Let us consider an  operator  ${\bf H} \in {\cal L}_{\rm symp}$:
\[{\bf H}= \left( \begin{array}{ll}
R&T\\
-T&R
\end{array}
\right )
\]
This operator defines the quadratic  Hamiltonian function
${\cal H}(q, p)=\frac{1}{2} ({\bf H} \om,\om)$
which can be written as
\begin{equation}
\label{HE1}{\cal H}(q, p)=\frac{1}{2}[(R p, p) + 2 (Tp, q) + (Rq, q)]
\end{equation}
where
$R^*=R , \; \; T^*=-T$
Corresponding Hamiltonian equations have the form
\begin{equation}
\label{HE2}
\dot q=Rp-Tq, \; \;  \dot p=-(Rq + Tp)
\end{equation}

{\bf Proposition 2.2.} {\it For a  symplectically invariant Hamilton function, the Hamiltonian flow $U_t,$ see (\ref{Y2}),  is $s$-commuting.}

\medskip

{\bf Example 2.1.} (One dimensional harmonic oscillator)
Let ${\cal H}(q, p)=\frac{1}{2} [\frac{p^2}{m} + m k^2 q^2]$
(we use the symbol $k$ to denote frequency, since $\omega$ is already used
for the point of the phase space).
 To get a Hamiltonian of the form (\ref{HE1}), we consider the case $\frac{1}{m}=m k^2.$ Thus
$\label{HE3}m=\frac{1}{k}$
and
${\cal H}(q, p)=\frac{k}{2} [ p^2 + q^2] ;$
Hamiltonian equations are given by
$ \dot q=k p,  \; \;  \dot p =-k q$
Here the symmetric $s$-commuting matrix
$${\bf H}= \left( \begin{array}{ll}
k&0\\
0&k
\end{array}
\right)
$$

{\bf 2.4. Symplectic form.} Let us define the {\it symplectic form} on the phase space:
$
w(\om_1, \om_2)=(\om_1, J\om_2)
$
Thus
$$
w(\om_1, \om_2)=(p_2, q_1)-(p_1, q_2)
$$
for $\omega_j= \{ q_j, p_j \}, j=1,2.$
This is a {\it skew-symmetric bilinear form.}

{\bf Proposition 2.3.} {\it Let $A$ be a symmetric operator. Then $A \in {\cal L}_{\rm symp}(\Omega)$ iff it is symmeric with respect to the symplectic form:}
\begin{equation}
\label{0}
w(A \om_1, \om_2)=w(\om_1, A\om_2)
\end{equation}

{\bf 2.5. Complex representation of dynamics for symplectically invariant Hamilton functions.}
Let us introduce on phase space $\Omega$ the complex structure: $\Om=Q \oplus i P$
We have $i\om=-p + iq=- J\om.$ A $\br$-linear operator $A:\Om \to \Om$ is ${\bf C}$-linear iff $A(i\om)=iA\om$ that is equivalent to $A \in {\cal L}_{\rm symp}.$

\medskip

{\bf Proposition 2.4.} {\it The class of ${\bf C}$-linear operators ${\cal L} ({\bf C}^{n})$ coincides with
the class of $s$-commuting operators ${\cal L}_{\rm symp}({\bf R}^{2n}).$}

\medskip

We introduce on $\Om$ a complex scalar product based on the ${\bf C}$-extension of the real scalar product:
$$
<\om_1, \om_2>=<q_1 + ip_1, q_2 + ip_2>
$$
$$
=(q_1, q_2) + (p_1, p_2) + i((p_1, q_2) - (p_2, q_1)).
$$
Thus
$
<\om_1, \om_2>=(\om_1, \om_2)-i w(\om_1, \om_2).
$

A ${\bf C}$-linear operator A  is symmetric with respect to the complex scalar product
$<\ldots>$ iff it is symmetric with respect to both real bilinear forms: ($\cdot, \cdot$) and w$(\cdot, \cdot)$.
Since for $A \in  {{\cal L}_{\rm symp}}$ the former implies the latter,  we get that a ${\bf C}$-linear operator is symmetric iff it is symmetric in the real space.

\medskip

{\bf Proposition 2.5.} {\it The class of ${\bf C}$-linear symmetric operators ${\cal L}_{\rm{s}}({\bf C}^{n})$ coincides with the class of $s$-commuting symmetric operators ${\cal L}_{\rm symp,s}({\bf R}^{2n}).$}

\medskip

We also remark that for a $s$-commuting operator $A$ its real and complex adjoint operators, $A^\star$ and $A^*,$
coincide. We showed that ${\bf C}$-linear symmetric operators appear naturally as complex
representations of $s$-commuting symmetric operators.

\medskip

{\bf Proposition 2.6.} {\it For a quadratic symplectically invariant Hamilton function the complexification does not change dynamics.}

\medskip

To prove this, we remark that $w( {\bf H} \om, \om)=0$ and hence
$$
{\cal H}(\om)= \frac{1}{2} <{\bf H} \om, \om> =\frac{1}{2} [({\bf H}\om, \om)- i w({\bf H}\om, \om)]=
\frac{1}{2} ({\bf H}\om, \om) , \om \in \Om.
$$
I consider complexification as merely using of new language: instead of symplectic invariance, we speak about ${\bf C}$-linearity. By Proposition 2.6 the Hamilton function (\ref{HE1}) can be written
${\cal H}(\omega)=\frac{1}{2} <{\bf H}\om, \om>, {\bf H} \in {\cal L}_{\rm{s}}({\bf C}^{n}),$
and  the correspoding  Hamiltonian equations can be written in the complex form as:
\begin{equation}
\label{Y10}
i \frac{d \om}{d t} = {\bf H} \om
\end{equation}
Any solution has the following complex representation:
\begin{equation}
\label{Y20}
\om(t)=U_t \om, \; \; U_t=e^{-i{\bf H} t}.
\end{equation}
This is the complex representation of flows corresponding to quadratic symplectically invariant
Hamilton functions.

\section{Schr\"odinger dynamics as a dynamics with symplectically invariant Hamilton function on
the infinite dimensional phase space}
Let $\Om\equiv H_c$ be a complex Hilbert space (infinite dimensional and separable) and let $<\cdot, \cdot>$ be the complex scalar product on $\Om.$
The symbol ${\cal L}_{\rm{s}}\equiv {\cal L}_{\rm{s}}(H_c)$ denotes the space of continuous ${\bf C}$-linear self-adjoint operators. The Schr\"odinger dynamics in $\Om$ is given by
\begin{equation}
\label{S} i h  \frac{d \om}{d t} = {\bf H} \om
\end{equation}
and hence
\begin{equation}
\label{S0}
\om(t)=U_t \om, \; \; U_t=e^{-i{\bf H}t/h}.
\end{equation}
We see that these are simply infinite-dimensional versions of equations (\ref{Y10}) and $(\ref{Y20})$ obtained from the Hamiltonian equations for quadratic symplectically invariant Hamilton function  in the process of complexification of classical  mechanics. Therefore we can reverse our previous considerations (with the only remark that
now the phase space is infinite dimensional) and represent the Schr\"odinger dynamics (\ref{S}) in the complex Hilbert space as the Hamiltonian dynamics in the infinite-dimensional phase space\footnote{Infinite dimension induces merely mathematical difficulties. The physical interpretation of formalism is the same as in the finite-dimensional case.}.
We emphasize that this Hamiltonian dynamics (\ref{Y1}) is a dynamics in the phase space $\Omega$ and not in the unit sphere of this Hilbert space! The Hamiltonian flow $\psi(t, \omega)=U_t\omega$ is a flow on the whole phase space $\Omega.$

We consider in $\Om$ the ${\bf{R}}$-linear operator $J$ corresponding to multiplication by $-i;$ we represent the complex Hilbert space in the form:
$$
\Om=Q \oplus i P,
$$
where $Q$ and $P$ are real Hilbert spaces:
$Q=P=H.$ As in the finite dimensional case, we have:

\medskip

{\bf Proposition 3.1.} {\it The class of continuous ${\bf C}$-linear self-adjoint operators ${\cal L}_{\rm{s}}(H_c)$ coincides with the class of  continuous $s$-commuting self-adjoint operators ${\cal L}_{\rm symp,s}(H \times H).$}

\medskip

Let us consider a quantum Hamiltonian ${\bf H} \in {\cal L}_{\rm{s}}$.\footnote{We may consider operator ${\bf H} \geq 0,$ but for the present consideration this is not important.} It is the image of the  classical Hamiltonian function. At the moment we operate only with quadratic physical classical variables. To find the quadratic form ${\cal H}(\omega)$ corresponding to ${\bf H},$ we should inverse the quantization map, see [] and section 8:
\[{\cal H}(\om)=\frac{1}{2h}  <{\bf H} \om, \om>=\frac{1}{2h} [(Rp, p) + 2(Tp, q)+(Rq, q)]\]
The corresponding Hamiltonian equation on the classical phase space $\Om=Q\times P,$ where $Q$ and $P$ are copies of the real Hilbert space is given by
\begin{equation}
\label{Z}
h\dot q=Rp-Tq, h\dot p=-(Rq + Tp)
\end{equation}
If we apply the complexification procedure to this system of Hamiltonian equations we, of course, obtain the Schr\"odinger equation (\ref{S}).

One may justify consideration of symplectically invariant physical variables on the Hilbert phase space
by referring to quantum mechanics: ``the correct classical Hamiltonian dynamics is based on symplectically invariant Hamilton functions, because they induce the correct quantum dynamics.'' So the classical prequantum dynamics
was reconstructed on the basis of the quantum dynamics. I have nothing against such an approach. But it would be interesting to find
internal classical motivation for considering symplectically invariant Hamilton functions. We shall do this in section 5.

\section{Lifting of point wise dynamics to spaces of variables and measures}

{\bf 4.1. General dynamical framework.}
Let $(X, F)$ be an arbitrary measurable space. So $X$ is a set and $F$ is a $\sigma$-field of its subsets.
Denote the space of random variables (measurable maps $f: X\to {\bf R})$ by the symbol $RV(X)$ and the
space of probability measures on $(X, F)$ by the symbol $PM(X).$ Consider a measurable map
$g:X \to X.$ It induces maps
$$
\alpha_g: RV(X) \to RV(X), \alpha_g f(x) =f(g(x))
$$
$$
\beta_g: MP(X) \to MP(X), \int_X f(x) d\beta_g \mu(x)  = \int_X \alpha_g f(x) d \mu(x).
$$
Now consider a dynamical system in $X:$
\begin{equation}
\label{PW0}
x_t= g_t(x),
\end{equation}
 where $g_t:X \to X$ is an one-parametric family of maps
(the parameter $t$ is real and plays the role of time). By using lifting $\alpha$ and $\beta$ we can lift
this point wise dynamics in $X$ to dynamics in $RV(X)$ and $MP(X),$  respectively:
\begin{equation}
\label{PW}f_t= \alpha_{g_t} f
\end{equation}
\begin{equation}
\label{PW1}\mu_t= \beta_{g_t} \mu.
\end{equation}
We shall see in sections 6,7 that for $X= \Omega$ (infinite dimensional phase space)  quantum images of dynamical systems
(\ref{PW0}), (\ref{PW}), (\ref{PW1}) are respectively dynamics of Schr\"odinger (for state -- wave function), Heisenberg (for operators-observables) and von Neumann (for density operator). To obtain quantum mechanics, we should choose adequate spaces of physical variables and measures.

{\bf 4.2. Lifting of the Hamiltonian dynamics.} It is well known that the lifting of Hamiltonian dynamics to the space of smooth
variables is given by the {\it Liouville equation}, see e.g. [2]. In particular, the functional lifting of any Hamiltonian dynamics on the Hilbert phase space $\Omega$ can be represented through the infinite-dimensional Liouville equation, [3]. We remark that
this is a general fact which has no relation to our special classical framework based on symplectically invariant Hamilton functions. For smooth functions on the $\Omega$ we introduce the Poisson brackets, see, e.g., [4]:
$$
\{ f_1(\omega), f_2(\omega)\}= \Big(\frac{\partial f_1}{\partial q}(\omega), \frac{\partial f_2}{\partial p}(\omega)\Big) -
\Big(\frac{\partial f_2}{\partial q}(\omega), \frac{\partial f_1}{\partial p}(\omega)\Big).
$$
We recall that for $f:H \to {\bf R}$ its first derivative can be represented by a vector belonging to  $H;$
so for $f: H \times H\to {\bf R}$ its gradient $\nabla f(\omega)$ belongs to $H \times H.$ We pay attention that $\{f_1, f_2\}=
=(\nabla f_1, J\nabla f_2)= w(\nabla f_1, \nabla f_2).$ Let ${\cal H}(\omega)$ be a smooth Hamilton function inducing the flow $\psi(t, \omega) =U_t(\omega).$  For a smooth function $f_0$ we set $f(t, \omega)= f_0(\psi(t, \omega)).$ It is
easy to see that this function is the solution of the Cauchy problem for the Liouville equation:
\begin{equation}
\label{PWZ}\frac{\partial f}{\partial t}(t,\omega)= \{ f(t, \omega), {\cal H} (\omega)\}, \; f(0, \omega)=f_0(\omega)
\end{equation}
The functional flow $\Psi(t,f_0)= \alpha_{U_t} f_0$ can be represented as
\begin{equation}
\label{PWZA}
\Psi(t,f_0)= e^{-t L} f_0,
\end{equation}
where
$$
L=\Big(\frac{\partial {\cal H} }{\partial q}(\omega), \frac{\partial }{\partial p}\Big) -
\Big(\frac{\partial {\cal H} }{\partial p}(\omega), \frac{\partial }{\partial q} \Big)
$$

\section{Dispersion preserving dynamics of statistical states}

Let us consider an arbitrary quadratic Hamiltonian function ${\cal H}(\omega)= \frac{1}{2} ({\bf H} \omega,  \omega)$
on the Hilbert phase $\Omega$ (the operator ${\bf H}$ need not be $s$-commuting). Let us consider
the Hamiltonian flow $U_t:  \Omega \to \Omega$ induced by the Hamiltonian system (\ref{Y}). This map is given by
(\ref{Y2}). It is important to pay attention that the map $U_t$ is invertible; in particular,
\begin{equation}
\label{PWO}U_t(\Omega)= \Omega.
\end{equation}We are interested in a Hamiltonian flow $U_t$ such that the corresponding dynamics in the space of probabilities
(\ref{PW1}) preserves magnitude of statistical fluctuations:
\begin{equation}
\label{PWD}
\sigma^2(\beta_{U_t}\mu) =\sigma^2(\mu):
\int_{\Omega} \Vert \omega \Vert^2 d\beta_{U_t}\mu (\omega) =
\int_{\Omega} \Vert \omega \Vert^2 d\mu (\omega)
\end{equation}
or
\begin{equation}
\label{PWD0}
\int_{\Omega} \Vert U_t\omega \Vert^2 d\mu (\omega)=
\int_{\Omega} \Vert \omega \Vert^2 d\mu (\omega) .
\end{equation}
Sufficient condition for preserving the magnitude of statistical fluctuations is preserving the magnitude of
individual fluctuations, i.e., the norm preserving:
\begin{equation}
\label{PWN}
\Vert U_t \omega \Vert^2 = \Vert \omega \Vert^2, \omega \in \Omega.
\end{equation}

{\bf Proposition 5.1.} {\it  The Hamiltonian flow corresponding to a quadratic Hamilton function ${\cal H}(\omega)$ is norm preserving iff the function ${\cal H}$ is symplectically invariant.}

{\bf Proof.} a). Let ${\cal H}$ be $s$-commuting. Then we have:
$$
\frac{d}{d t} \Vert U_t \omega\Vert^2= 2  (\dot{U_t}\omega, U_t \omega) = 2(J {\bf H} U_t \omega, U_t \omega)=0
$$
Here we used the simple fact that the operator $J{\bf H}$ is  skew symmetric: $(J {\bf H})^\star= - {\bf H}J= - J{\bf H}.$
Thus (\ref{PWN}) holds.

b). Let  (\ref{PWN}) hold. Then $\frac{d}{d t} \Vert U_t \omega \Vert^2=0.$ By using previous computations and (\ref{PWO}) we get
that the operator $J{\bf H}$ is  skew symmetric. This implies that ${\bf H}$ commutes with $J.$

\medskip

In particular, {\it only the Hamiltonian flow corresponding to a symplectically invariant Hamilton function preserves the fluctuations of the Planck magnitude.} This is our explanation of the exceptional role of symplectically invariant physical variables on the
infinite-dimensional classical phase space.

If a Hamilton function is not symplectically invariant then the corresponding
Hamiltonian flow can induce increasing of the magnitude of fluctuations. But we recall that quantum model is a representation based on neglecting by fluctuations of the magnitude $o(h), h \to 0.$ Therefore  a
Hamiltonian flow which is not  symplectically invariant  can induce the
transformation of ``quantum statistical states'', i.e., distributions on the phase space having dispersion of the magnitude $h,$
into ``nonquantum statistical states'', i.e. distributions on the phase space having dispersions essentially larger
than $h.$

\section{Dynamics in the space of quadratic symplectically invariant physical variables}

{\bf 6.1. Lifting of Hamiltonian dynamics to the space of quadratic variables.}
Let us consider the  Hamiltonian flow $U_t:  \Omega \to \Omega$ induced by an arbitrary  quadratic Hamilton function.
Let $A:\Omega \to \Omega$ be a continuous self-adjoint operator and  $f_A=(A\omega, \omega).$
We have $\alpha_{U_t} f_A(\omega)= f_A(U_t \omega)= f_{U_t^\star A U_t}(\omega).$
This dynamics can be represented as the dynamics in the space of continuous linear symmetric operators
\begin{equation}
\label{DH}A_t= U_t^\star A U_t
\end{equation}
We remark that $U_t= e^{J{\bf H}t/h},$ so $U_t^\star= e^{-{\bf H}J t/h}.$ Thus
\begin{equation}
\label{DHZ}A_t= e^{-{\bf H}J t/h} A e^{J{\bf H}t/h}.
\end{equation}Thus
$\frac{d A_t}{d t} = \frac{1}{h} \Big( A_t J{\bf H} - {\bf H}J A_t\Big),$ or
\begin{equation}
\label{DH0}
\frac{d A_t}{d t} = \frac{1}{h}[A_t, {\bf H}J]+ \frac{1}{h} A_t [J, {\bf H}]
\end{equation}
We remark that dynamics (\ref{DH}) can be also obtained from the Liouville equation, but I presented the direct derivation.

{\bf 6.2. Lifting for symplectically invariant variables.}
We consider the space of physical variables
$$
V_{\rm{quad, symp}}(\Omega)=\{ f: \Omega \to {\bf R}: f \equiv f_A(\omega) = \frac{1}{2} (A \omega, \omega),
A \in {\cal L}_{\rm{symp}, s}(\Omega) \}
$$
(consisting of symplectically invariant quadratic forms).  Let us consider the lifting of the  flow corresponding
to a symplectically invariant quadratic Hamilton function to the space $V_{\rm{quad, symp}}(\Omega).$ In this case
both operators, ${\bf H}$ and $A$ are $s$-commuting. Therefore the flow (\ref{DHZ}) can be written as
\begin{equation}
\label{DHZQ}A_t= U_t^\star A U_t = e^{-J {\bf H} t/h} A e^{J{\bf H}t/h}
\end{equation}
The evolution equation (\ref{DH0}) is simplified:
\begin{equation}
\label{DH0A}
\frac{d A_t}{d t} = \frac{-J}{h}[ {\bf H},A_t]
\end{equation}

{\bf 6.3. Complexification.} By considering on the phase space the complex structure and representing the symplectic operator  $J$ by $-i$ we write (\ref{DHZ}) in the form of the Heisenberg dynamics:
\begin{equation}
\label{DHG}
A_t= U_t^* A U_t=e^{it{\bf H}/h} A e^{-it{\bf H}/h}
\end{equation}
(here $U_t^*$ is the complex adjoint operator to $U_t)$
and the evolution equation (\ref{DH0}) in the form of the Heisenberg equation:
\begin{equation}
\label{DH0G}
\frac{d A_t}{d t} = \frac{i}{h}[ {\bf H},A_t]
\end{equation}
Thus this equation is just the image of the lifting of the  classical quadratic Hamiltonian dynamics  in the case of symplectically invariant variables.

\section{Dynamics in the space of Gaussian distributions}

{\bf 7.1. Lifting of  Hamiltonian dynamics to the space of Gaussian measures.} Let us consider a  Hamiltonian flow $U_t:  \Omega \to \Omega$ induced by an arbitrary  quadratic Hamilton function.  Let $\rho$ be an arbitrary Gaussian measure with zero mean value. Since  a linear continuous transformation of a Gaussian measure is
again a Gaussian measure, we have that $\beta_{U_t}(\rho)$ is Gaussian. We find dynamics of the covariation operator of $\beta_{U_t}(\rho).$ We have:
$$
(\rm{cov}(\beta_{U_t}\rho) y_1, y_2)=   \int_\Om  (y_1, \omega) (y_2, \omega) d \beta_{U_t} \rho (\omega)
$$
$$
= \int_\Om  (y_1, U_t \omega) (y_2, U_t\omega) d  \rho (\omega)= (\rm{cov} (\rho) U_t^\star y_1, U_t^\star y_2).
$$
Thus, for the covariation operator $B_t= \rm{cov}(\beta_{U_t}\rho),$ we have:
\begin{equation}
\label{DH0M}
B_t= U_t B U_t^\star \equiv e^{J{\bf H} t/h} B e^{-{\bf H}J t/h}
\end{equation}
Thus
$\frac{d B_t}{d t} = \frac{1}{h} \Big( J{\bf H}B_t  -  B_t{\bf H}J\Big),$ or
\begin{equation}
\label{DHM}
\frac{d B_t}{d t} = \frac{1}{h}[J{\bf H}, B_t]+ \frac{1}{h} B_t [J, {\bf H}]
\end{equation}

{\bf 7.2. Lifting for symplectically invariant measures.} We now consider the lifting of the flow
induced by symplectically invariant quadratic Hamilton function. We start with the following mathematical result:

{\bf Proposition 7.1.} {\it A Gaussian measure (with zero mean value) is symplectically invariant if  its covariation operator is symplectically invariant.}

{\bf Proof.} a). Let $\rho$ be a Gaussian measure with zero mean value and $B=\rm{cov}(\rho).$ Let $\beta_J \rho= \rho.$
It is sufficient to prove that $B J$ is skew  symmetric. We have:
$$
(BJy_1, y_2)=
= \int_\Omega (Jy_1, \omega) (y_2, \omega) d \rho(\omega)=
- \int_\Omega (y_1, J\omega) (y_2,  J^\star J \omega) d \rho(\omega)
$$
$$
=- \int_\Omega (y_1, J\omega) (Jy_2,   J \omega) d \rho(\omega)=
- \int_\Omega (y_1, \omega) (Jy_2,   \omega) d \beta_J \rho(\omega)
$$
$$
=
- \int_\Omega (Jy_2,   \omega) (y_1, \omega) d \rho(\omega)=
-(BJy_2,y_1).
$$
b). Let $B=\rm{cov}(\rho)\in {\cal L}_{\rm{symp}, s}(\Omega).$ We find the Fourier transform
of the Gaussian measure $\beta_J \rho:$
$$
\widetilde{\beta_J \rho}(y)= \int_\Omega e^{i(y, J\omega)} d \rho (\omega)=
=\tilde{\rho}(J^\star y)= e^{-\frac{1}{2} (B J^\star y, J^\star y)}= =\tilde{\rho}(y).
$$

From the proof we also obtain:

{\bf Corollary 7.1.} {\it Let $\rho$ be an arbitrary symplectically invariant measure. Then
its covariation operator is symplectically invariant.}

Since the flow for a symplectically invariant (quadratic) Hamilton function is $s$-commuting, by using the representation
(\ref{DH0M}) and Proposition 7.1 we prove that the space of symplectically invariant Gaussian measures (with zero mean value) is invariant for the map $\beta_{U_t}.$  Here we have:
\begin{equation}
\label{DH0MA}
B_t= U_t B U_t^\star \equiv e^{J{\bf H}t/h} B e^{-J{\bf H} t/h}
\end{equation}
 or
\begin{equation}
\label{DHMA}
\frac{d B_t}{d t} = \frac{-J}{h}[B_t, {\bf H}]
\end{equation}

{\bf 7.3. Complexification.} By considering on the phase space the complex structure and representing the symplectic operator  $J$ by $-i$ we write (\ref{DH0MA}) in the form:
\begin{equation}
\label{DH0MB}
B_t=U_t B U_t^*=  e^{-i{\bf H}t/h} B e^{i{\bf H} t/h}
\end{equation}
 or
\begin{equation}
\label{DHMX}
\frac{d B_t}{d t} = \frac{i}{h}[B_t, {\bf H}]
\end{equation}
This is nothing else than the von Neumann equation for the statistical operator. The only difference is that
the covariance operator $B$ is not normalized. The normalization will come from the correspondence map $T$
projecting a prequantum classical statistical model onto QM, see section 8.

{\bf 7.4. Dynamics in the space of statistical states.}
First we consider the space of all Gaussian measures having zero mean value and dispersion $2h.$ Denote it by the symbol  $S_G^h(\Omega).$ These are Gaussian measures such that
$$
(y, m_\rho)= \int_\Omega (y, \omega) d\rho(\omega)=0, y \in \Omega, \; \mbox{and}\;  \sigma^2(\rho)= \int_\Omega \Vert \omega\Vert^2 d \rho(\omega)= 2h
$$
{\bf Remark 7.1} We choose fluctuations having dispersion $\sigma^2(\rho)=2h$ to obtain ``pure states''
corresponding to fluctuations with covariation matrices (which are of the size $2\times2)$
having eigenvalues $\lambda_1=\lambda_2 =h.$ So in that case $\sigma^2(\rho)=2h= \rm{Tr}\; B= h+h,$
see section 9 for more details.

For the flow $U_t$ corresponding to a symplectically invariant quadratic Hamilton function, we have (see section 5)
$
\beta_{U_t}: S_G^h(\Omega) \to S_G^h(\Omega)
$
Denote the subspace of $S_G^h(\Omega)$ consisting of symplectically invariant measures by the symbol $S_{G, \rm{symp}}^h(\Omega).$ We also have:
$$
\beta_{U_t}: S_{G, \rm{symp}}^h(\Omega) \to S_{G, \rm{symp}}^h(\Omega).
$$

{\bf 7.5. Complex covariation.} Everywhere below we consider only measures with finite dispersions. Let us introduce  {\it complex average and covariance operator},  $m_\rho^c$ and $B^c\equiv \rm{cov}^c \rho,$ by setting:
\begin{equation}
\label{CCAV}
<m_\rho^c, y>=  \int_\Omega <y, \omega> d \rho (\omega).
\end{equation}
\begin{equation}
\label{CCV} <B^c y_1, y_2>= \int_\Omega <y_1, \omega> <\omega, y_2> d \rho (\omega).
\end{equation}

{\bf Proposition 7.2.} {\it Let $\rho$ be a  symplectically invariant measure. Then}
\begin{equation}
\label{GSAV}m_\rho^c =0 \; \mbox{iff}\; \;  m_\rho=0.
\end{equation}
{\bf Proof.} Since $\rho$ is symplectically invariant, for any Borel function $f: \Omega \to {\bf R},$
we have:
\begin{equation}
\label{GSA0}\int_\Omega f(\omega_q,\omega_p) d\rho(\omega_q,\omega_p)= \int_\Omega f(\omega_p, -\omega_q) d\rho(\omega_q,\omega_p)
\end{equation}
Let $m_\rho=0.$ Then:
$$
0=\int_\Omega (y, \omega) d\rho(\omega) = \int_\Omega [(y_q, \omega_q) + (y_p, \omega_p)] d\rho(\omega)
$$
$$
=\int_\Omega [(y_q, \omega_p) - (y_p, \omega_q)] d\rho(\omega) =  \int_\Omega w(y, \omega)d\rho(\omega).
$$
Hence the last integral is also equal to zero. On the other hand, for the complex average we  have:
\begin{equation}
\label{GSAC}
<y, m_\rho^c> =0= \int_\Omega (y, \omega) d\rho(\omega) - i \int_\Omega w(y, \omega)d\rho(\omega).
\end{equation}

{\bf Proposition 7.3.} {\it Let $\rho$ be a  symplectically invariant measure. Then}
\begin{equation}
\label{GSI}\rm{cov}^c \rho = 2 \rm{cov} \rho
\end{equation}

{\bf Proof.} a We have
$$
\rm{cov}^c \rho (y, y)= \int_\Om \vert < y, \om>\vert^2 d\rho(\omega)=
\int_\Om \vert (y, \omega) - i w(y, \om)\vert^2 d\rho(\omega)
$$
$$
= \int_\Om [(y, \omega)^2 + (y, J\om)^2] d\rho(\omega).
$$
By using symplectic invariance of the measure $\rho$ we get:
$$
\int_\Om (y, J\om)^2 d\rho(\omega)=
\int_\Om (y, \om)^2 d\rho(\omega).
$$
Thus
$$
\rm{cov}^c \rho (y, y)= 2 \int_\Om (y, \om)^2 d\rho(\omega)=
2 \rm{cov} \rho (y, y).
$$

\medskip

{\bf Theorem 7.1.} {\it For any measure $\rho$ and $s$-commuting operator $A$, we have:
\begin{equation}
\label{GSIZ}
\int_\Omega (A\omega, \omega) d\rho(\omega) = \rm{Tr} \; \rm{cov}^c \rho \; A;
\end{equation}
in particular, }
\begin{equation}
\label{GSI0}
\sigma^2(\rho) = \rm{Tr}\;  \rm{cov}^c \rho.
\end{equation}

{\bf Proof.} Let $\{e_j\}$ be an orthonormal basis in $H_c$ (we emphasize that orthogonality and
normalization are with respect to the complex and not real scalar product). Then:
$$
\rm{Tr}\;  \rm{cov}^c \rho\; A =
 \int_\Om \sum_j  < Ae_j, \om> <\om, e_j> d\rho(\omega)=
\int_\Omega <A\omega, \omega> d\rho(\omega)
$$
$$
 = \int_\Omega (A\omega, \omega) d\rho(\omega)
$$

\medskip

We recall that we showed in  [1] that
and
\begin{equation}
\label{GSI1}
\sigma^2(\rho) = \rm{Tr}\; \rm{cov} \rho.
\end{equation}

It seems that there is a contradiction between equalities (\ref{GSI1}), (\ref{GSI0}) and (\ref{GSI}). In fact, there is no contradiction, because in (\ref{GSI1}) and (\ref{GSI0}) we use two different  traces: with respect to the real and complex  scalar products, respectively. This  is  an important point; even normalization by trace one for the von Neumann density operator is the normalization with respect to the complex scalar product.

We remark that  the complex average $m_\rho^c$ and the covariation operator $B^c$  are ${\bf C}$-linear even if a measure is not symplectically invariant. However, in general real and complex averages do not coincide and real and complex
covariance operators are not coupled by (\ref{GSI}).

Let us find relation between $B={\rm cov} \rho$ and $B^c={\rm cov}^c \rho$ in the general case. It is easy to see that for
\[ B=\left( \begin{array}{ll}
B_{11} & B_{12}\\
B_{21} & B_{22}
\end{array}
\right ), B^*_{11}=B_{11}, B_{22}^*=B_{22}, B_{12}^*=B_{21}
\]
and
\[ B^c=\left( \begin{array}{ll}
D & S\\
-S & D
\end{array}
\right )
\]
we have

{\bf Proposition 7.4.}
{\it The blocks in real and complex covariation operators are connected by the following equalities:}
\begin{equation}
\label{COM}
D=B_{11} + B_{22}, S=B_{12}-B_{21}.
\end{equation}

Thus {\it in the general case the complex covariation operator $B^c$ does not determine the Gaussian measure $\rho_B$ uniquely.}

Let now $\rho_B$ be symplectically invariant. Then
\[ B=\left( \begin{array}{ll}
B_{11} & B_{12}\\
-B_{12} & B_{11}
\end{array}
\right ).
\]
Thus
\begin{equation}
\label{COM1}
D=2 B_{11}, S= 2 B_{12},
\end{equation}
so we obtain (\ref{GSI}) and, hence, we obtain:

\medskip

{\bf Corollary 7.2.} {\it There is one-to-one correspondence between symplectically invariant Gaussian measures and complex covariation operators.}\footnote{These are ${\bf C}$-linear self-adjoint positively defined operators $B^c: H_c \to H_c$ belonging to the trace class}.

\section{Prequantum classical statistical model}

{\bf 8.1. Quadratic variables.} We consider the classical  statistical model
$$
M_{\rm{quad}}= ( S_{G, \rm{symp}}^h(\Omega), V_{\rm{quad, symp}}(\Omega)),
$$
where $\Om= Q\times P$ and $Q=P=H,$ and the conventional (Dirac-von Neumann)
quantum statistical model $N_{rm{quant}}=( {\cal  D}(H_c),
{\cal L}_{\rm{s}}(H_c)),$ where ${\cal  D}(H_c)$ is the space of density operators and
${\cal L}_{\rm{s}}(H_c)$ is the space of bounded self-adjoint operators in $H_c$
(quantum observables).\footnote{To simplify considerations, we consider only quantum observables
represented by bounded operators. To obtain the general quantum model with
observables represented by unbounded operators, we should consider a prequantum classical
statistical model based on the Gelfand triple: $H_c^{+} \subset H_c \subset H_c^{-}.$}

The classical $\to$ quantum correspondence map $T$ is similar to the map presented in [1]
for the real case:
\begin{equation}
\label{TT1}T: S_{G, \rm{symp}}^h(\Omega) \to {\cal  D}(H_c),
\; \; T(\rho)= \frac{\rm{cov}^c \rho}{2h}
\end{equation}
\begin{equation}
\label{TT2}T: V_{\rm{quad, symp}}(\Omega) \to {\cal L}_{\rm{s}}(H_c),
\; \; T(f)=  h f^{\prime \prime}(0)
\end{equation}

\medskip

{\bf Theorem 8.1.} (On properties of the classical $\to$ quantum correspondence) {\it The map $T$  is one-to-one on the spaces $S_{G, \rm{symp}}^h(\Omega)$ and  $V_{\rm{quad, symp}}(\Omega);$ the map $T: V_{\rm{quad}}(\Omega) \to {\cal L}_{s}(H_c)$ is ${\bf R}$-linear and the fundamental equality of classical and quantum averages holds:}
\begin{equation}
\label{TT3}<f>_\rho= \int_\Om f(\om) d \rho(\om)=
\rm{Tr}\; T(\rho) T(f)= \rm{Tr} \; \rm{cov}^c  \rho \; f^{\prime \prime}(0).
\end{equation}

\medskip
This equality is the consequence of Theorem 7.1.

{\bf 8.2. Analytic variables.} As in the case of the real Hilbert space $H,$ see [1], we can extend essentially
the class of variables.  Let us consider, cf. [1], the functional space ${\cal V}_{\rm{symp}}(\Omega)$
consisting of real analytic functions, $f:\Omega \to {\bf R},$ which have the exponential growth:
\begin{equation}
\label{COR1}\mbox{there exist}\;  C, \alpha \geq 0 :  \vert f(\omega)\vert \leq C e^{\alpha \Vert x\Vert};
\end{equation}
preserve the state of vacuum:
\begin{equation}
\label{COR2}
f(0)=0
\end{equation}
and which are symplectically invariant.

The following trivial mathematical result plays the fundamental role in establishing classical $\to$ quantum correspondence.

{\bf Proposition 8.1.} {\it Let $f \in {\cal V}_{\rm{symp}}(\Omega).$ Then }
\begin{equation}
\label{COR3}
 f^{\prime \prime}(0)\in {\cal L}_{\rm{symp}, s}(\Omega).
\end{equation}

\medskip

We consider now the classical statistical model:

\begin{equation}
\label{MH}
M_{\rm{a,symp}}= ( S_{G, \rm{symp}}^h(\Omega),{\cal V}_{\rm{symp}}(\Omega)).
\end{equation}
The classical $\to$ quantum correspondence map $T$ is defined by (\ref{TT1}), (\ref{TT2}). However, the equality of averages (\ref{TT3}) is, of course, violated, cf. [1].
Let us find the average of a variable $f \in {\cal V}_{\rm{symp}}(\Omega)$ with respect to a statistical state $\rho_B \in S_{G,\rm{symp}}^h(\Omega):$
$$
<f>_{\rho_B}= \int_\Omega f(\omega) d\rho_B(\omega) = \int_\Omega f(\sqrt{2h} \omega^\prime) d\rho_D (\omega^\prime)
$$
\begin{equation}
\label{ANN1}
=\sum_{n=2}^\infty  \frac{(2h)^{n/2}}{n!} \int_\Omega f^{n}(0)(\omega^\prime, ...,\omega^\prime)d\rho_D (\omega^\prime),
\end{equation}
where the covariation operator of the scaling transformation $\rho_D$ of the Gaussian measure $\rho_B$ has the form:
$$
D=B/2h.
$$
Since $\rho_B\in S_G^h(H),$ we have $\rm{Tr} \; D  = 1.$
The change of variables in (\ref{ANN1}) can be considered as rescaling of the magnitude of
statistical  (Gaussian) fluctuations. Fluctuations which were considered as very small,
\begin{equation}
\label{DS2}
\sigma^2 (\rho)= 2 h,
\end{equation}
(where $h$ is a small parameter) are considered in the new scale as standard normal fluctuations.\footnote{Thus QM is a kind of the statistical microscop which gives us the possibility to see the effect of fluctuations of  the Planck magnitude (near the  the vacuum field $\omega=0)$.}
By (\ref{ANN1}) we have:
\begin{equation}
\label{ANN2}
<f>_\rho=  h  \int_\Omega (f^{\prime \prime}(0)\omega^\prime, \omega^\prime) d\rho_D(\omega^\prime) + o(h), \; h \to 0,
\end{equation}
or
\begin{equation}
\label{ANN3}<f>_\rho =  h  \rm{Tr}\; D \; f^{\prime \prime}(0) + o(h), \; h \to 0.
\end{equation}
We see that the classical average (computed in the model (\ref{MH}) by using measure-theoretic approach)
is approximately equal to the quantum average (computed in the model
$N_{\rm{quant}}=({\cal D}(H_c), {\cal L}_{\rm{s}}(H_c))$ with the aid of the von Neumann trace-formula).

\medskip

{\bf Theorem 8.2.} {\it  For the classical statistical model $M_{\rm{a,symp}}$ the map $T,$ see (\ref{TT1}),
(\ref{TT2}), performing classical $\to$ quantum correspondence is one-to-one on the space of statistical states
$S_{G, \rm{symp}}^h(\Omega),$ but it has a huge degeneration on the space of physical variables
${\cal V}_{\rm{symp}}(\Omega).$ Classical and quantum averages are in general not equal, but the asymptotic equality
(\ref{ANN3}) holds.}

\medskip

{\bf Remark 8.1.} (Magnitude of prequantum Gaussian fluctuations) We considered statistical states given by
(symplectically invariant) Gaussian measures with dispersion $\sigma^2(\rho) = 2h.$ From the physical point of view
it is more natural to consider statistical states with dispersions:
\begin{equation}
\label{Inna}
\sigma^2(\rho) = 2h + o(h), h \to 0.
\end{equation}
The only difference is that by projecting such a classical statistical model on the quantum model $N_{\rm{quant}}$
we shall not obtain one-to-one correspondence between classical and quantum statistical states.

{\bf References}

[1] A. Yu. Khrennikov, Prequantum classical statistical model with infinite dimensional phase-space.
submited to {\it J. Phys. A: Math. General.}

[2] N. N. Bogolubov and N. N Bogolubov (son), {\it Introduction to quantum statistical mechanics.}
Nauka (Fizmatlit): Moscow (1984).

[3] A.Yu. Khrennikov, Infinite-Dimensional equation of Liuville.
{\it Mat. Sbornik}, {\bf 183}, 20-44 (1992).

[4] A. Yu. Khrennikov, The principle of correspondence in quantum
theories of field and relativistics bosonic string. {\it Mat.
Sbornic}, {\bf 180}, 763-786 (1989);  {\it Supernalysis.} Nauka, Fizmatlit, Moscow, 1997 (in
Russian). English translation: Kluwer, Dordreht, 1999.

[5]  J. von Neumann,  {\it  Mathematical foundations of quantum mechanics.} Princeton Univ. Press: Princeton, N.J. (1955).

[6] A. S. Holevo, {\it Statistical structure of quantum theory,} Springer,
Berlin-Heidelberg (2001).

\end{document}